\documentstyle[11pt]{article}

\begin{document}

\setcounter{page}{1}
\pagestyle{plain} \vspace{1cm}
\begin{center}
\Large{\bf Realization of blue spectrum in generalized Galileon super-inflation models}\\
\small \vspace{1cm} {\bf K. Nozari$^{a,}$\footnote{knozari@umz.ac.ir}}\quad and \quad {\bf S. Shafizadeh$^{b,}$\footnote{s.shafizadeh@tpnu.ac.ir}}\\
\vspace{0.5cm} {\it $^{a}$Department of Physics, Faculty of Basic
Sciences,\\
University of Mazandaran,\\
P. O. Box 47416-95447, Babolsar, IRAN\\
{\it $^{b}$ Department of Physics, Payame Noor University (PNU),\\
P. O. Box 19395-3697, Tehran, Iran}}

\begin{abstract}
In the spirit of Galileon inflation and by considering some sorts of non-canonical kinetic terms in the action,
we realize a stage of super-inflation leading to a blue-tilted tensor perturbation. We show also that
addition of Galileon-like term to the action leads to avoidance of ghost instabilities in this setup.\\

\end{abstract}

\end{center}

\vspace{1cm}
\section{INTRODUCTION}

An important outcome of the inflationary scenario for very early
stage of the universe evolution [1] is that it yields hypothetical
framework for generation and growth of the scalar and tensor
perturbations [2]. Such a framework can be tested via observations
such as the CMB anisotropies, and could be used in essence as a tool
to determine viability of different models of inflation [3].

Although many inflationary models predict observable gravitational
wave background, primordial tensor perturbations (gravitational
waves) construct a little contribution in the temperature
perturbations (anisotropies) of the cosmic background radiation.
Measurements of the CMB polarization can help in essence to detect
these perturbations. The CMB polarization decomposes to orthogonal
components, one of them is E-mode (curl-free) that is generated by
density perturbations at recombination, hence is related to
temperature anisotropies of the CMB radiation. Another one, the
B-mode, is divergence-free and originates from the differential
stretching of spacetime related to a background of initial
gravitational waves [4]. Fortunately and finally the gravitational
wave has been detected by the LIGO Scientific Collaboration and
Virgo Collaboration in a context other than the CMB anisotropy
(in the context of Binary Black Hole Merger) [5]. In this regard, we need some more theoretical
advancements to see the role of tensor perturbation in cosmological structure
formation. Since the tensor power spectrum depends only on the
Hubble expansion rate during inflation, these power spectrum is
think carefully as a straight investigation of the scale of
inflation. In this regard, it is very often deduced that the tensor
spectrum from vacuum fluctuations is always red-tilted. On the other
hand, in standard inflation models, the hubble expansion rate reduces in a
gradual way $(\dot{H}<0)$ and involves $\epsilon>0$ where
$\epsilon$ is the slow-roll parameter. Therefore, we can conclude
that the tensor spectral index of perturbations is always negative
and leads to fluctuation modes with shorter wavelengths having lower
magnitudes than those fluctuations with longer wavelength. Although
the recent observational data are unable to determine the spectrum
of tensor perturbation in CMB polarization, the possibility of being
blue-tilted for inflationary perturbations is corresponding to the
negative values of $\epsilon$. This condition implies violation of
the null energy condition that is concluded from $\dot{H}>0$
[6,7,8].

This feature can be encoded in the notion of super-inflation. Loop
quantum cosmology as an approach to super-inflation predicts an era of super-inflation
by modified curvature sector of the Einstein-Hilbert action. These
quantum modifications cause to achieve super-inflation epoch which
happens during the early universe, independently of the shape of
potential [9,10,11].

As another approach, it is possible also to realize an epoch of
super-inflation via addition of non-canonical kinetic terms (in the
presence of Galileon term) in the Lagrangian. Here, we suggest a simple
mechanism to realize blue-tilted tensor spectrum in an era of
super-inflation, with $\dot{H}>0$, by considering some sorts of
non-canonical kinetic terms in the Lagrangian. In fact, one
advantage of super-inflation stage is that it can modify the
spectral tilt by making it towards blue-tilted one before the
conventional slow-roll inflation kicks in with a decreasing $H(t)$,
which is known to yield a red-tilted spectrum [12]. Prediction of
the super-inflation epoch shows that $H$ may grow during inflation when
the null energy condition, $\rho+P\geq0$ (with $\rho$ the energy
density and $P$ the pressure), is violated.

Based on these preliminaries, in this paper we concentrate on the
spectral index of primordial perturbations produced in a
super-inflation stage. Our idea is that perturbations could give
rise in a period of super-inflation during which the Hubble rate
rapidly grows, instead of staying almost constant as it is the case
throughout standard slow-roll inflation regime. Our purpose here is
to study the possibility of blue tilted primordial perturbations in
super-inflation regime by considering models of inflation containing
running kinetic terms and non-canonical kinetic term in the action
and in the spirit of G-inflation (see [13-17] for G-inflation). By
using higher order kinetic terms we provide required conditions for
rapidly growing of Hubble expansion parameter and creation of blue
spectrum of perturbations without need to large contribution of
potential of the scalar field. We show that incorporation of the
Galileon-like term in this setup, which modifies the k-inflation,
results in avoidance of ghost instability.

\section{G-Inflation and background equations}
We concentrate on a generalized G-inflation model with a scalar
field. We phenomenologically consider a dimensionless scalar field
$\varphi$ with a action of the type
\begin{equation}
\label{eq1}S=\sum^{4}_{i=2}\int d^{4}x\sqrt{-g}{\cal{L}}_{i}\,.
\end{equation}
Here $g$ is the determinant of the metric $g_{\mu\nu}$ and [18]
\begin{equation}
\label{eq2}{\cal{L}}_{2}=K(\phi,X)\,,
\end{equation}
\begin{equation}
\label{eq3}{\cal{L}}_{3}=-G_{3}(\phi,X)\Box\phi\,,
\end{equation}
\begin{equation}
\label{eq4}{\cal{L}}_{4}=G_{4}R+G_{4X}\Bigg[(\Box\phi)^{2}-(\nabla_{\mu}\nabla_{\nu}\phi)^{2}\Bigg]\,,
\end{equation}
where $R$ is the Ricci tensor, $K$ and $G_{i}$ are arbitrary
functions of $\phi$ and
$X=-(\frac{1}{2})g^{\mu\nu}\nabla_{\mu}\phi\nabla_{\nu}\phi$,
$G_{i}(\phi,X)=g_{i}(\phi)+h_{i}(\phi)X$ and $\Box\equiv
g^{\mu\nu}\nabla_{\mu}\nabla_{\nu}$ is the standard d'Alembertian
operator. This theory was originally found by Horndeski [16,19] in a
different form. In the recent form, it can be used as a framework to
study the single-field inflation model. This G-inflation model
contains k-inflation, the non-minimal coupling between the scalar
field and gravity and also the non-minimal derivative coupling to
the Einstein tensor. In this setup we remove higher derivatives that
would in other respects appear in the field equations, and consider
the G-inflation which includes the second derivatives of the scalar
field. We consider inflation model with the following choices
\begin{equation}
\label{eq5}K(X,\phi)={\cal{K}}(\phi)X-V\,,
\end{equation}
\begin{equation}
\label{eq6}G_{3}=-\gamma(\phi)X\,,
\end{equation}
\begin{equation}
\label{eq7}G_{4}=\frac{1}{2}(m^{2}_{pl}+\xi\phi^{2})+\frac{1}{2\mu^{2}}X\,,
\end{equation}
where $\gamma(\phi)$ is a dimensionless function of the scalar
field. We first derive the background equations for the theories
given by the action $(1)$ on the flat Friedmann-Robertson-Walker
(FRW) geometry with the scalar factor $a(t)$ of the form
$ds^{2}=-dt^{2}+a^{2}(t)\delta_{ij}dx^{i}dx^{j}$ where $t$ is the cosmic
time. Taking variation of the action at first order with respect to
the metric, leads to the Friedmann equations as
\begin{equation}
\label{eq8}H^{2}=\frac{\rho_{\phi}}{3m^{2}_{pl}}\,,\,\,\,\,\,\
\dot{H}=-\frac{\rho_{\phi}+P_{\phi}}{2m^{2}_{pl}} \,,
\end{equation}
where the energy density and the pressure of the
scalar field correspondingly can be written as follows:
\begin{equation}
\label{eq9}\rho_{\phi}=\frac{1}{2}\dot{\phi}^{2}\Bigg[{\cal{K}}(\phi)-6\gamma\
H\dot{\phi}+\gamma_{\phi}\dot{\phi}^{2}+\frac{9}{\mu^{2}}
H^{2}-12\xi\frac{H\phi}{\dot{\phi}}-6\xi\frac{H^{2}\phi^{2}}{\dot{\phi}^{2}}\Bigg]
+V \,,
\end{equation}
\begin{eqnarray}
\label{eq10}p_{\phi}=\frac{1}{2}\dot{\phi}^{2}\Bigg[{\cal{K}}(\phi)+\gamma_{\phi}
\dot{\phi}^{2}+2\gamma
\ddot{\phi}-4\frac{H\ddot{\phi}}
{\mu^{2}\dot{\phi}}-\frac{3H^{2}+2\dot{H}}{\mu^{2}}+4\xi\\\nonumber
+2\bigg(\frac{m^{2}_{pl}+\xi\phi^{2}}{\dot{\phi}^{2}}-1\bigg)
\Big(3H^{2}+2\dot{H}\Big)
+4\xi\phi\bigg(\frac{\ddot{\phi}}{\dot{\phi}^{2}}+\frac{2H}{\dot{\phi}}\bigg)\Bigg]-V
\end{eqnarray}
The scalar field equation of motion is given by variation of the action
(1) with respect to $\phi(t)$
\begin{equation}
\label{eq11}\frac{1}{a^{3}}\frac{d}{dt}(a^{3}J)={\cal{P}}_{\phi}
\end{equation}
where
\begin{equation}
\label{eq12}J=\dot{\phi}\Big({\cal{K}}(\phi)+\frac{3}{\mu^{2}}H^{2}\Big)-3\gamma
H\dot{\phi}^{2}+\gamma_{\phi}\dot{\phi}^{3}
\end{equation}
and
\begin{equation}
\label{eq13}{\cal{P}}_{\phi}=\frac{1}{2}{\cal{K}}_{\phi}\dot{\phi}^{2}-V_{\phi}+
\frac{1}{2}\dot{\phi}^{2}\bigg(2\gamma_{\phi}\ddot{\phi}+
\gamma_{\phi\phi}\dot{\phi}^{2}+12\frac{\xi\phi}{\dot{\phi}^{2}}\Big(2H^{2}+\dot{H}\Big)
\bigg)
\end{equation}
Finally, by substituting equations (12) and (13) into the
equation (11) we can derive the following equation of motion
\begin{eqnarray}
\label{eq14}\ddot{\phi}\Big({\cal{K}}-6\gamma
H\dot{\phi}+\frac{3}{\mu^{2}}H^{2}+2\gamma_{\phi}\dot{\phi}^{2}\Big)
+3H\dot{\phi}\Big({\cal{K}}-3\gamma
H\dot{\phi}+\frac{3}{\mu^{2}}H^{2}\Big)\\\nonumber
+\frac{1}{2}\dot{\phi}^{2}\Big({\cal{K}}_{\phi}-6\gamma\dot{H}+
\gamma_{\phi\phi}\dot{\phi^{2}}\Big)
-6\xi\phi\Big(2H^{2}+\dot{H}\Big)+V_{\phi}=0
\end{eqnarray}
Hence, we can derive the scalar field equation of state parameter as
\begin{eqnarray}
\label{eq15}w_{\phi}\equiv\frac{p_{\phi}}{\rho_{\phi}}=
\frac{\frac{1}{2}\dot{\phi}^{2}\Big[{\cal{K}}+\gamma_{\phi}
\dot{\phi}^{2}+2\gamma \ddot{\phi}-4(\frac{H\ddot{\phi}}
{\mu^{2}\dot{\phi}})+4\xi
+2\big(\frac{(m^{2}_{pl}+\xi\phi^{2})}{\dot{\phi}^{2}}-1-\frac{1}{2\mu^{2}}\big)
(3H^{2}+2\dot{H}) +4\xi\phi\Big(\frac{\ddot{\phi}}{\dot{\phi}^{2}}+
\frac{2H}{\dot{\phi}}\Big)\Big] -V}
{\dot{\phi}^{2}\Big[{\cal{K}}(\phi)-6\gamma\
H\dot{\phi}+\gamma_{\phi}\dot{\phi}^{2}+\frac{9}{\mu^{2}}
H^{2}-12\xi\frac{H\phi}{\dot{\phi}}-6\xi\frac{H^{2}\phi^{2}}{\dot{\phi}^{2}}\Big]
+V}
\end{eqnarray}
and the deceleration parameter as
\begin{equation}
\label{eq16}q\equiv-1-\frac{\dot{H}}{H^{2}}=\frac{1}{2}+\frac{3}{2}w_{\phi}\,.
\end{equation}
Rapid roll condition during the super-inflation results in the following relation

\begin{equation}
\label{eq17}\frac{\dot{\phi}}{H}\simeq
-(m^{2}_{pl}+\xi\phi^{2})\frac{F}{V\bigg({\cal{K}}+\frac{1}{\mu^{2}}U+
\sqrt{({\cal{K}}+\frac{1}{\mu^{2}}U)^{2}+4\gamma F}\bigg)}
\end{equation}
where
\begin{equation}
\label{eq18}F=\frac{V_{\phi}}{m^{2}_{pl}+\xi(1+6\xi)\phi^{2}}\,\,,\,\,\,\,\,\,\, U=\frac{V}{m^{2}_{pl}+\xi\phi^{2}}\,.
\end{equation}

\section{Quadratic action for tensor and scalar perturbations}

In this section we study tensor and scalar cosmological
perturbations in generalized G-inflation by action expansion method.
Let us first consider unitary gauge in which $\delta\phi=0$ and
begin by writing the perturbed metric in the Arnowitt-Deser-Misner
(ADM) formalism as $[20]$
\begin{equation}
\label{eq19}ds^{2}=-N^{2}dt^{2}+h_{ij}(dx^{i}+N^{i}dt)(dx^{j}+N^{j}dt)
\end{equation}
where $N$ and $N^{i}$ are the lapse and shift functions
\begin{equation}
\label{eq20}N=1+2\Phi,\,\,\,N_{i}=\delta_{ij}\partial^{j}B\,, \,\,\nonumber
h_{ij}=a^{2}(t)e^{2\Psi}\delta_{ij}
\end{equation}
Here, $\Phi$, $\Psi$ and $B$ are scalar perturbations and $h_{ij}$
is defined as tensor perturbation. Now we write the perturbed metric
at the linear level as $[21,22,23]$
\begin{equation}
\label{eq21}ds^{2}=-(1+2\Phi)dt^{2}+2a^{2}(t)B_{,i}dx^{i}dt+a^{2}(t)(1-2\Psi)\delta_{ij}dx^{i}dx^{j}
\end{equation}
Expanding the action $(1)$ up to the second order in the perturbations,
we obtain the following result
\begin{eqnarray}
\label{eq22}S_{2}=\int
dtd^{3}xa^{3}\Bigg\{-3\Big((m^{2}_{pl}+\xi\phi^{2})-\frac{X}{2\mu^{2}}\Big)\dot{\Psi}^{2}
+\frac{1}{a^{2}}\bigg[2\Big((m^{2}_{pl}+\xi\phi^{2})-\frac{X}{2\mu^{2}}\Big)\dot{\Psi}-\bigg(\frac{6}{\mu^{2}}H
X+2H(m^{2}_{pl}+\xi\phi^{2})\\\nonumber+2\dot{\phi}(\gamma
X+\xi\phi\big)\bigg)\Phi\bigg]\partial^{2}B-\frac{2}{a^{2}}\Big((m^{2}_{pl}+\xi\phi^{2})-\frac{X}{2\mu^{2}}\Big)
\Phi\partial^{2}\Psi+\bigg(\frac{2}{\mu^{2}}H
X+H(m^{2}_{pl}+\xi\phi^{2})\\\nonumber+\dot{\phi}(\gamma
X+\xi\phi\big)\bigg)\Phi\dot{\Psi}+\bigg[X\Big({\cal{K}}-12\gamma\dot{\phi}H+
\frac{18}{\mu^{2}}\Big)+4\gamma_{\varphi}X^{2}-3H^{2}(m^{2}_{pl}+\xi\phi^{2})-
6H\xi\phi\dot{\phi}\bigg]\Phi^{2}+\\\nonumber
\frac{1}{a^{2}}\Big((m^{2}_{pl}+\xi\phi^{2})+\frac{X}{\mu^{2}}\Big)(\partial\Psi)^{2}\Bigg\}\,.
\end{eqnarray}
While the coefficients are quite complicated, the construction of
the action (22) is similar to that expressed in Ref. [24]. By using
the above second order action, we can find the momentum and
Hamiltonian constrains as
\begin{equation}
\label{eq23}\Phi=\frac{(m^{2}_{pl}+\xi\phi^{2})-\frac{X}{\mu^{2}}}{\bigg(-\frac{3}{\mu^{2}}H
X+H(m^{2}_{pl}+\xi\phi^{2})\\\nonumber+\dot{\phi}(\gamma
X+\xi\phi\big)\bigg)}\dot{\Psi}
\end{equation}
\begin{eqnarray}
\label{eq24}\frac{1}{a^{2}}\partial^{2}B=\frac{X\Big({\cal{K}}-12\gamma\dot{\phi}H+
\frac{18}{\mu^{2}}\Big)+4\gamma_{\phi}X^{2}-3H^{2}(m^{2}_{pl}+\xi\phi^{2})-6H\xi\phi\dot{\phi}}
{\frac{3}{\mu^{2}}H X+H(m^{2}_{pl}+\xi\phi^{2})+\dot{\phi}(\gamma
X+\xi\phi\big)}\Phi\\\nonumber
+3\dot{\Psi}-\frac{1}{a^{2}}\Bigg(\frac{2(m^{2}_{pl}+\xi\phi^{2})-\frac{X}{\mu^{2}}}
{\frac{6}{\mu^{2}}H X+2H(m^{2}_{pl}+\xi\phi^{2})+2\dot{\phi}(\gamma
X+\xi\phi\big)}\Bigg)\partial^{2}\Psi,\,\
\end{eqnarray}
By substituting Eq. (23) into Eq. (22), the second order
action reduces to the following form
\begin{equation}
\label{eq25}S_{2}=\int
dtd^{3}xa^{3}{\cal{U}}\Big[\dot{\Psi}^{2}-\frac{c_{s}^{2}}{a^{2}}(\partial\Psi)^{2}\Big]
\end{equation}
where
\begin{eqnarray}
\label{eq26}{\cal{U}}=\frac{\Big((m^{2}_{pl}+\xi\phi^{2})
-\frac{X}{\mu^{2}}\Big)^{2}\bigg[X\Big({\cal{K}}-12\gamma\dot{\phi}H+
\frac{18}{\mu^{2}}H^{2}+4\gamma_{\phi}X\Big)-3H^{2}(m^{2}_{pl}+\xi\phi^{2})-6H\xi\phi\dot{\phi}\bigg]}
{\Big[-\frac{3}{\mu^{2}}HX+H(m^{2}_{pl}+\xi\phi^{2})+\dot{\phi}(\gamma
X+\xi\phi)\Big]^{2}}\\
\nonumber+3\Big(m^{2}_{pl}+\xi\phi^{2}-\frac{X}{\mu^{2}}\Big)
\end{eqnarray}
and
\begin{equation}
\label{eq27}c^{2}_{s}=\frac{4-\frac{{\cal{K}}+\frac{V}{\frac{1}{\mu^{2}}(m^{2}_{pl}+\xi\phi^{2})}}
{{\cal{K}}+(\frac{1}{\mu^{2}})U+\sqrt{({\cal{K}}+(\frac{1}{\mu^{2}})U)^{2}+4\gamma
F}}}{3\Bigg[2-\frac{{\cal{K}}+\frac{V}{\frac{1}{\mu^{2}}(m^{2}_{pl}+\xi\phi^{2})}}
{{\cal{K}}+(\frac{1}{\mu^{2}})U+\sqrt{({\cal{K}}+(\frac{1}{\mu^{2}})U)^{2}+4\gamma
F}}\Bigg]}
\end{equation}
To avoid ghosts and Laplacian instabilities we require the following conditions
\begin{equation}
\label{eq28}{\cal{U}}>0,\,\,\,\,\,\,c^{2}_{s}\geq0
\end{equation}
The power spectrum of the curvature perturbations $\Psi$, is given
by
\begin{equation}
\label{eq29}{\cal{P}}_{\Psi}=\frac{H^{2}}{8\pi^{2}{\cal{U}}c^{3}_{s}}
\end{equation}

The second order action for the tensor perturbations in this model can be written
as
\begin{equation}
\label{eq30}S_{T}=\int
dtd^{3}xa^{3}{\cal{Q}}\big[\dot{h}^{2}_{ij}-\frac{c_{T}^{2}}{a^{2}}(\partial
h_{ij})^{2}\big]
\end{equation}
where
\begin{equation}
\label{eq31}{\cal{Q}}=\frac{(m^{2}_{pl}+\xi\phi^{2})-\frac{X}{2\mu^{2}}}{4}\,,
\end{equation}
\begin{equation}
\label{eq32}c^{2}_{T}=\frac{(m^{2}_{pl}+\xi\phi^{2})+\frac{X}{\mu^{2}}}{(m^{2}_{pl}+\xi\phi^{2})-\frac{X}{2\mu^{2}}}\,.
\end{equation}
The power spectrum of primordial tensor perturbations is given by
\begin{equation}
\label{eq33}{\cal{P}}_{T}=\frac{H^{2}}{2\pi^{2}{\cal{Q}}c^{3}_{T}}\,.
\end{equation}

\section{Applications to concrete models of Super-Inflation}

Now we consider parameter space of some specific models in this setup in order
to study the blue spectrum of tensor and scalar modes and also to
explore the conditions for super-inflation regime.

At first, let us apply the case in which the field has
non-canonical form. We start from the action of the type

\begin{equation}
\label{eq34}{\cal{L_{\phi}}}=K(\phi,X)\,,
\end{equation}
where $X\equiv-\frac{1}{2}\nabla_{\mu}\phi\nabla^{\nu}\phi$ and
\begin{equation}
\label{eq35}K(\phi,X)=-V(\phi)+{\cal{K}}(\phi)X\,,\, G_{3}=0\,,\,\,
G_{4}=\frac{1}{2}m^{2}_{pl}\,.
\end{equation}

The energy density and pressure of the scalar field
are written respectively as follows
\begin{equation}
\label{eq36}\rho_{\phi}={\cal{K}}(\phi)X+V(\phi),\,\,\,\,P_{\phi}={\cal{K}}(\phi)X-V(\phi)
\end{equation}
With these quantities, it is easy to check that increment of the
Hubble parameter with time during super-inflation stage yields the important result
$(\rho_{\phi}+P_{\phi})<0$ which is violation of the null energy
condition in this case. As a result, we are able to achieve ${\cal{K}}<0$ which means that
${\cal{K}}$ can be a negative function of the scalar field.
One important point should be stressed here: in Ref. [25] the authors have shown that in potential-dominant generalized G-inflation models
the blue tensor spectrum is prohibited in order to avoid instabilities. The assumptions made in our case seems to meet the postulates of Ref. [25]. However, this is not actually the case since the authors of [25] assume ${\cal{K}}(\phi)$ to be positive definite while here we consider the possibility of having negative ${\cal{K}}$ in super inflation stage. In fact, as has been shown in Refs. [7,10], in super inflation stage ${\cal{K}}$ can be negative. This is the main deference of our work with Ref. [25] in this respect.
We note also that, an
accurate description of universe during the phase of super-inflation
can be acquire with a fast rolling scalar field as $X\gg V(\phi)$.
In other words, the dominated contribution is related to kinetic
energy. In this case, the rapid roll condition can be provided so
that ${\cal{K}}X\ll V(\phi)$, with ${\cal{K}}$ that is negative
function of the scalar field. Also by taking time derivative of both sides
of this condition we obtain $V_{\phi}\gg({\cal{K_{\phi}}}X+{\cal{K}}\ddot{\phi})$.
Then we can derive the equation of motion for the scalar field as
\begin{equation}
\label{eq37}{\cal{K}}\ddot{\phi}+3{\cal{K}}H\dot{\phi}+{\cal{K}}_{\phi}X+V_{\phi}=0
\end{equation}
By considering the condition $V_{\phi}\gg
({\cal{K_{\phi}}}X+{\cal{K}}\ddot{\phi})$ the equation of motion is
given by
\begin{equation}
\label{eq38}3{\cal{K}}H\dot{\phi}+V_{\phi}\simeq0\,
\end{equation}
so we achieve the result
$\frac{\dot{\phi}}{H}\simeq-\frac{V_{\phi}}{{\cal{K}}V}$.
Then by using the second order action of the scalar and tensor
perturbations, we can express spectral indexes in this model in
terms of $\phi$-dependent functions as
\begin{equation}
\label{eq39}n_{T}\simeq-2\epsilon\simeq-\frac{V^{2}_{\phi}}{{\cal{K}}V}\,.
\end{equation}
When we consider ${\cal{K}}<0$, this leads to $n_{T}>0$. Hence the
tensor spectrum can be blue tilted in this model of super-inflation. Also for $n_s$ we find
\begin{equation}
\label{eq40}n_{s}-1\simeq-4\epsilon+\eta-\frac{\dot{J}}{HJ}=-\frac{3V^{2}_{\phi}}{{\cal{K}}V^{2}}
-\frac{V_{\phi}{\cal{K}}_{\phi}}{V{\cal{K}}^{2}}+\frac{2V_{\phi\phi}}{{\cal{K}}V}\,.
\end{equation}
For scalar modes, if $\epsilon<0$ that is provided
by ${\cal{K}}<0$, then $n_{s}>1$ is possible in this case. To proceed further,
we introduce ansatz for $V(\phi)$ and ${\cal{K}}(\phi)$ as
$V(\phi)=\phi^{n}$ and ${\cal{K}}(\phi)=-\phi^{1/n}$ with $n=2$.

\begin{figure}
\begin{center}\includegraphics{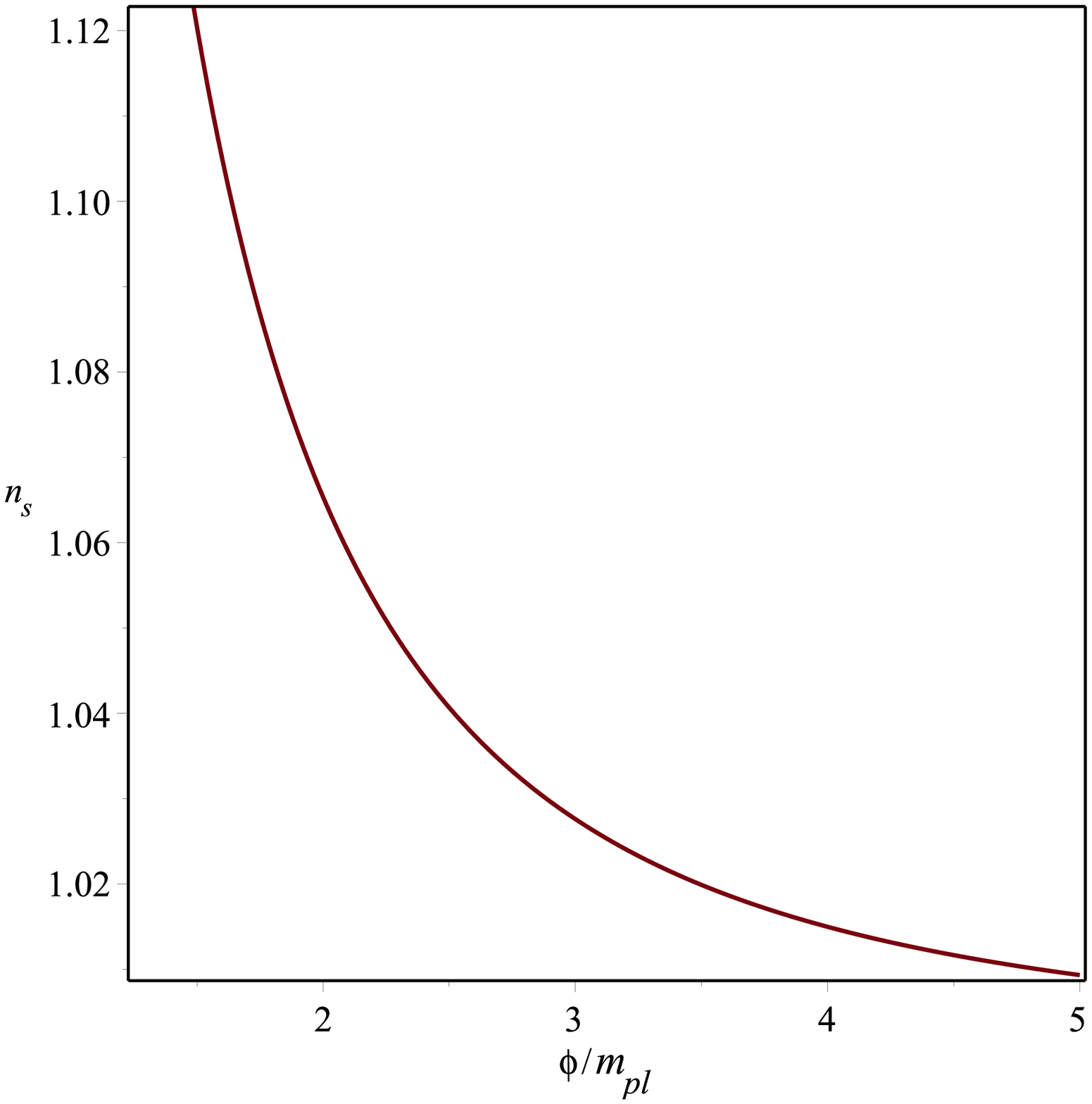} \vspace{7cm}\includegraphics{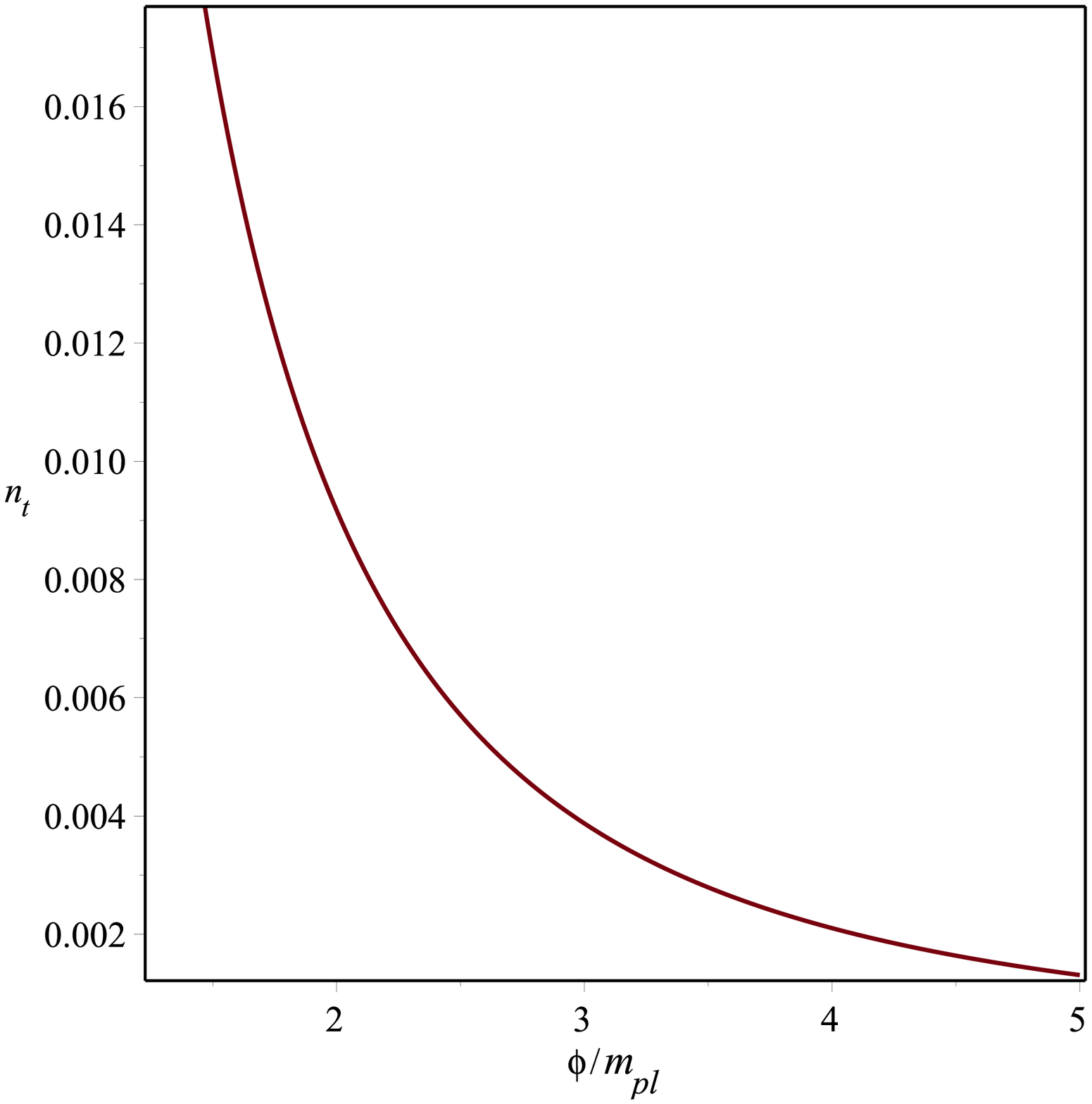}
\end{center}
\caption{\small {Variation of the scalar spectral index (left panel) and tensor spectral index
(right panel) versus $\phi/m_{pl}$  for quadratic potential.}}
\end{figure}

Figure 1 (right panel) shows the variation of the tensor spectral
index $n_{T}$ versus the scalar field. Obviously, in the presence of
running kinetic term, the tensor spectral index is positive. So, the
tensor spectrum can be blue tilted in super-inflation regime.
The left panel of this figure shows the variation of the scalar spectral index $n_{s}$ versus
scalar field. The spectral index can be larger than unity which shows the
possibility of having blue spectrum in this case.

Note that with rapid roll condition in super-inflation regime, it is
seen that $w_{\phi}<-1$. It shows $\epsilon<0$ and we also achieve
$0<c^{2}_{s}<1$. We expect a transition from the super-inflation
regime to slow roll inflation in the infrared (IR) regime. On the
other hand, increasing Hubble rate of expansion continues until
$\dot{H}\rightarrow0$. In fact, there is a reversal in the rate of
$H(t)$ then Hubble parameter starts to decrease in slow roll phase
of the universe because of slowly rolling of the scalar field, for a
review see [26] . It can be seen easily that
\begin{equation}
\label{eq41}\dot{H}\rightarrow0\,\, \Rightarrow\,\, \epsilon\rightarrow0\,\, \Rightarrow
n_{T}\rightarrow0
\end{equation}
and this condition holds if $V_{\phi}\rightarrow0$. Also, it leads to $n_{s}\rightarrow1$,
the scale invariance of the perturbation. We note that in this case there is ghost instability
since ${\cal{U}}<0$ as can be seen in figure 2. About the equation of state parameter, we have

\begin{equation}
\label{eq42}w_{\phi}=-1+\frac{V^{2}_{\phi}}{3{\cal{K}}V^{2}},\,\,\,\,\,
c^{2}_{s}\simeq1
\end{equation}
that is corresponding to the super-inflation regime. It means
$w_{\phi}<-1$ and also $\epsilon<0$ in this case.

\begin{figure}
\begin{center}\includegraphics{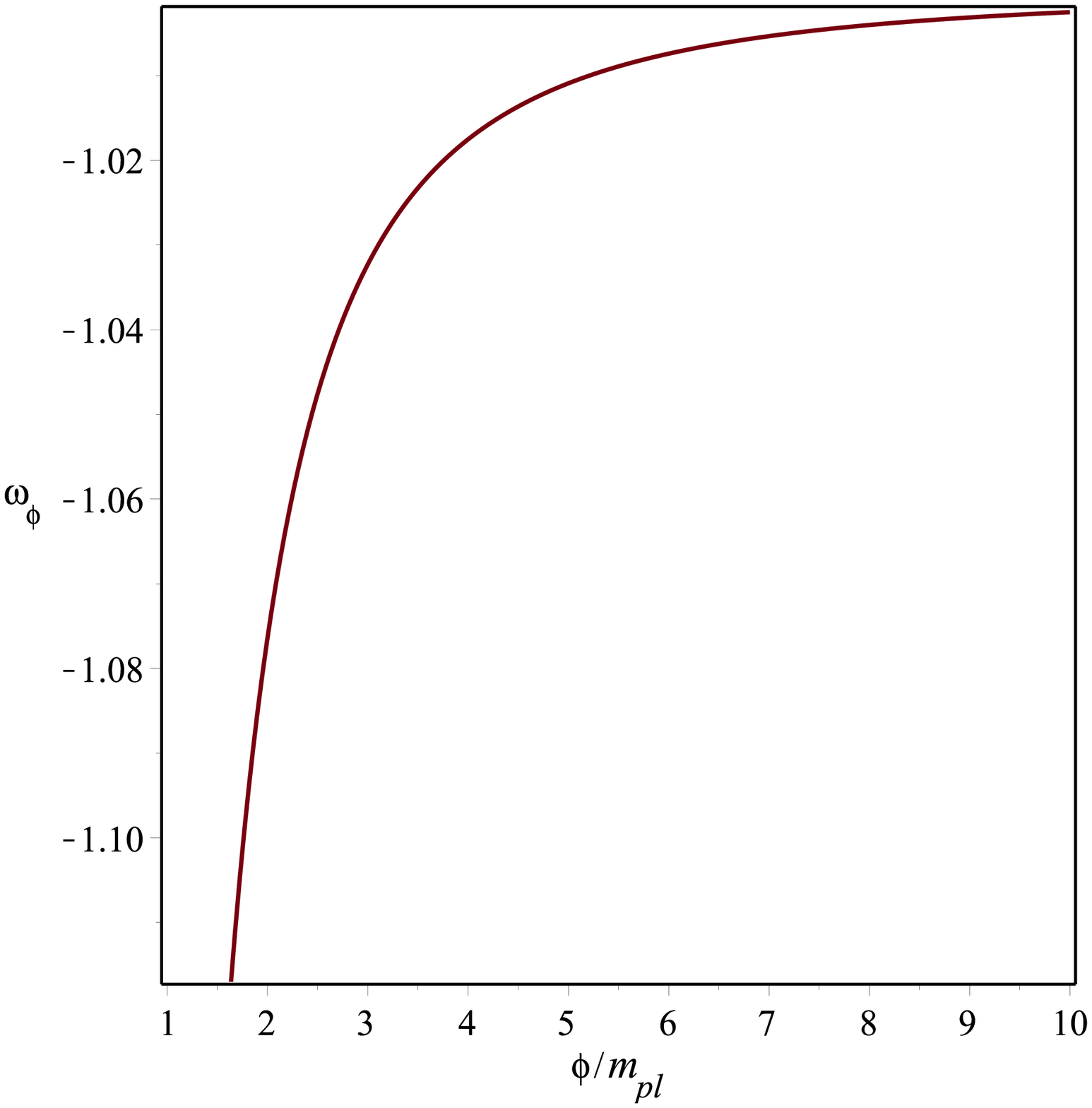} \vspace{7cm}\includegraphics{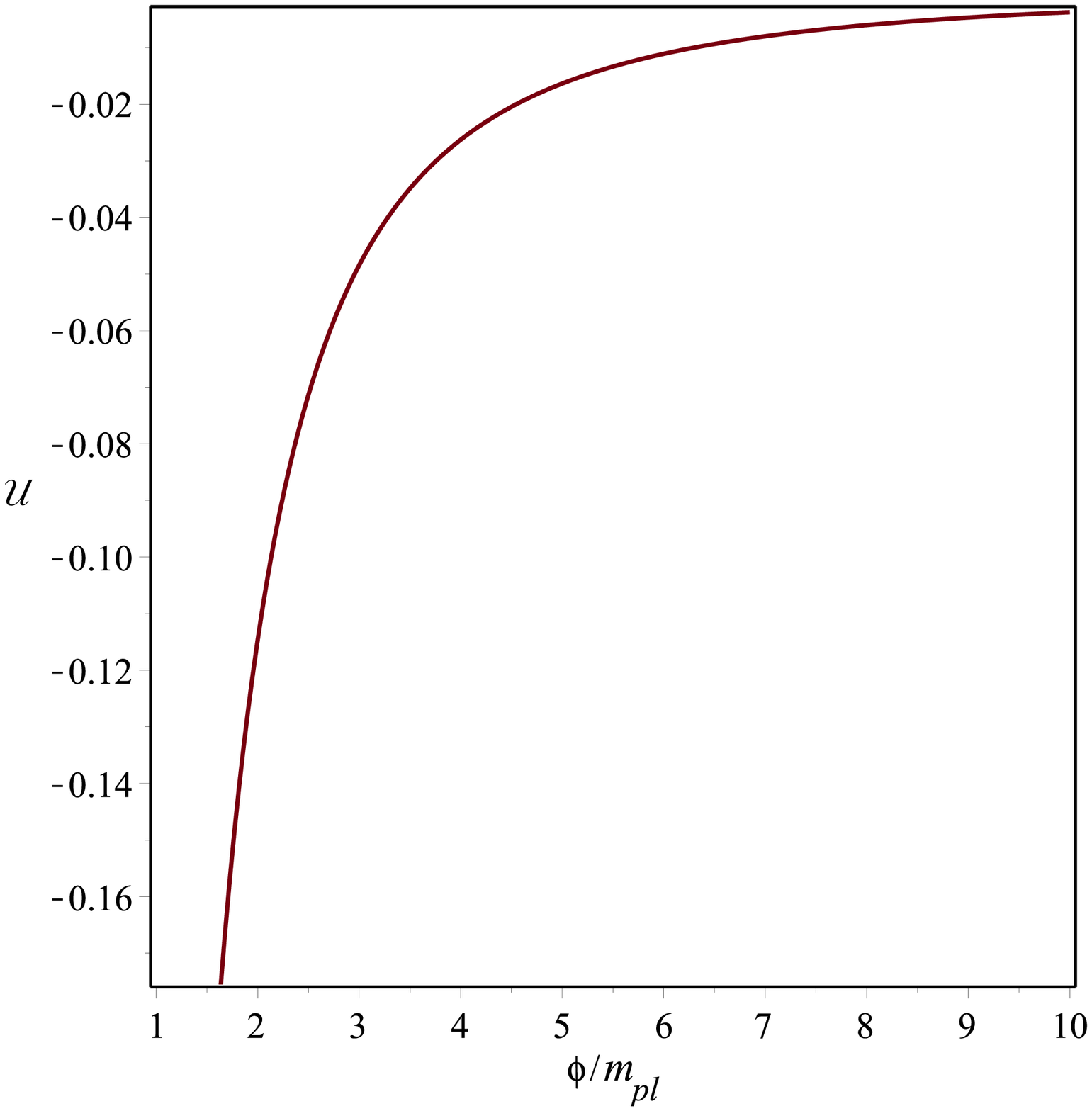}
\end{center}
\caption{\small {Variation of the equation of state parameter (left panel) and ghost instability
parameter $({\cal{U}})$ (right panel) versus $\phi/m_{pl}$  for
quadratic potential.}}
\end{figure}
Figure $2$ (left panel) shows the variation of the equation of state parameter versus
the scalar field. In this class of inflationary models
$w_{\phi}<-1$ but there is ghost instability in this setup (right panel of figure 2) as in k-inflation model.

In the next step, we consider a model with G-super inflation action in which it
contains Galileon-like kinetic term. We limit our attention to the
alternative to super inflation model with
\begin{equation}
\label{eq43}K(X,\varphi)={\cal{K}}(\phi)X-V\,,\,\,
G_{3}=-\gamma(\phi)X\,,\,\, G_{4}=\frac{1}{2}m^{2}_{pl}\,,
\end{equation}
where $\gamma(\varphi)$ is a  negative dimensionless function of the
scalar field. In this case Friedmann equations are written as follows
\begin{equation}
\label{eq44}3m^{2}_{pl}H^{2}=X\Bigg[{\cal{K}}(\phi)-6\gamma\
H\dot{\phi}+\gamma_{\phi}\dot{\phi}^{2}\Bigg] +V \,,
\end{equation}
\begin{eqnarray}
\label{eq45}\dot{H}=-X\Bigg[{\cal{K}}(\phi)-3\gamma
H\dot{\phi}+\gamma_{\phi}\dot{\phi}^{2}+\frac{1}{2}\gamma\ddot{\phi}\Bigg]
\end{eqnarray}
During the super-inflation phase, $\dot{H}>0$ which is
corresponding to $(\rho_{\phi}+P_{\phi})<0$ to have
fast roll expansion in this phase of the universe expansion.
Also the energy density needs to be positive.
Therefore it is required from equations (43)
and (44) together to have the following conditions
\begin{equation}
\label{eq46}{\cal{K}}X<V(\phi),\,\,\,\,\mid\gamma H
\dot{\phi}\mid<V(\phi),\,\,\,\, {\cal{K(\phi)}}\ll0 \,.
\end{equation}
Using these conditions in equation of motion, we find
\begin{equation}
\label{eq47}\frac{\dot{\phi}}{H}\simeq\frac{1}{2\gamma
V}\Bigg({\cal{K}}-\sqrt{{\cal{K}}^{2}+4\gamma V_{\phi}}\Bigg)
\end{equation}
Now, the spectral indexes of perturbations can be obtained as
\begin{equation}
\label{eq48}n_{T}\simeq\Big(\frac{V_{\phi}}{V}\Big)^{2}\Big(\frac{-2}{{\cal{K}}+\sqrt{{\cal{K}}^{2}+4\gamma
V_{\phi}}}\Big)
\end{equation}
and
\begin{equation}
\label{eq49}n_{s}\simeq1-\frac{V_{\phi}}{V[{\cal{K}}+\sqrt{{\cal{K}}^{2}+4\gamma
V_{\phi}}]}\Bigg[6\frac{V_{\phi}}{V}+\frac{\gamma_{\phi}}{\gamma}+3V_{\phi\phi}-
\Bigg(\frac{(\frac{{\cal{K}}}{{\cal{K}}+\sqrt{{\cal{K}}^{2}+4\gamma
V_{\phi}}})_{\phi}}{1-\frac{{\cal{K}}}{{\cal{K}}+\sqrt{{\cal{K}}^{2}+4\gamma
V_{\phi}}}}-\frac{3(\frac{{\cal{K}}}{{\cal{K}}+\sqrt{{\cal{K}}^{2}+4\gamma
V_{\phi}}})_{\phi}}{2-\frac{{\cal{K}}}{{\cal{K}}+\sqrt{{\cal{K}}^{2}+4\gamma
V_{\phi}}}}\Bigg)\Bigg]
\end{equation}
We consider the following functions

\begin{equation}
\label{eq50}{\cal{K}}=-\phi^{\frac{1}{4}},\,\,\,
\gamma=-\alpha\phi^{n},\,\,\,\ V(\phi)=\phi^{2}
\end{equation}
where $\alpha<1$ and we set $n=1$ in numerical analysis.

\begin{figure}
\begin{center}\includegraphics{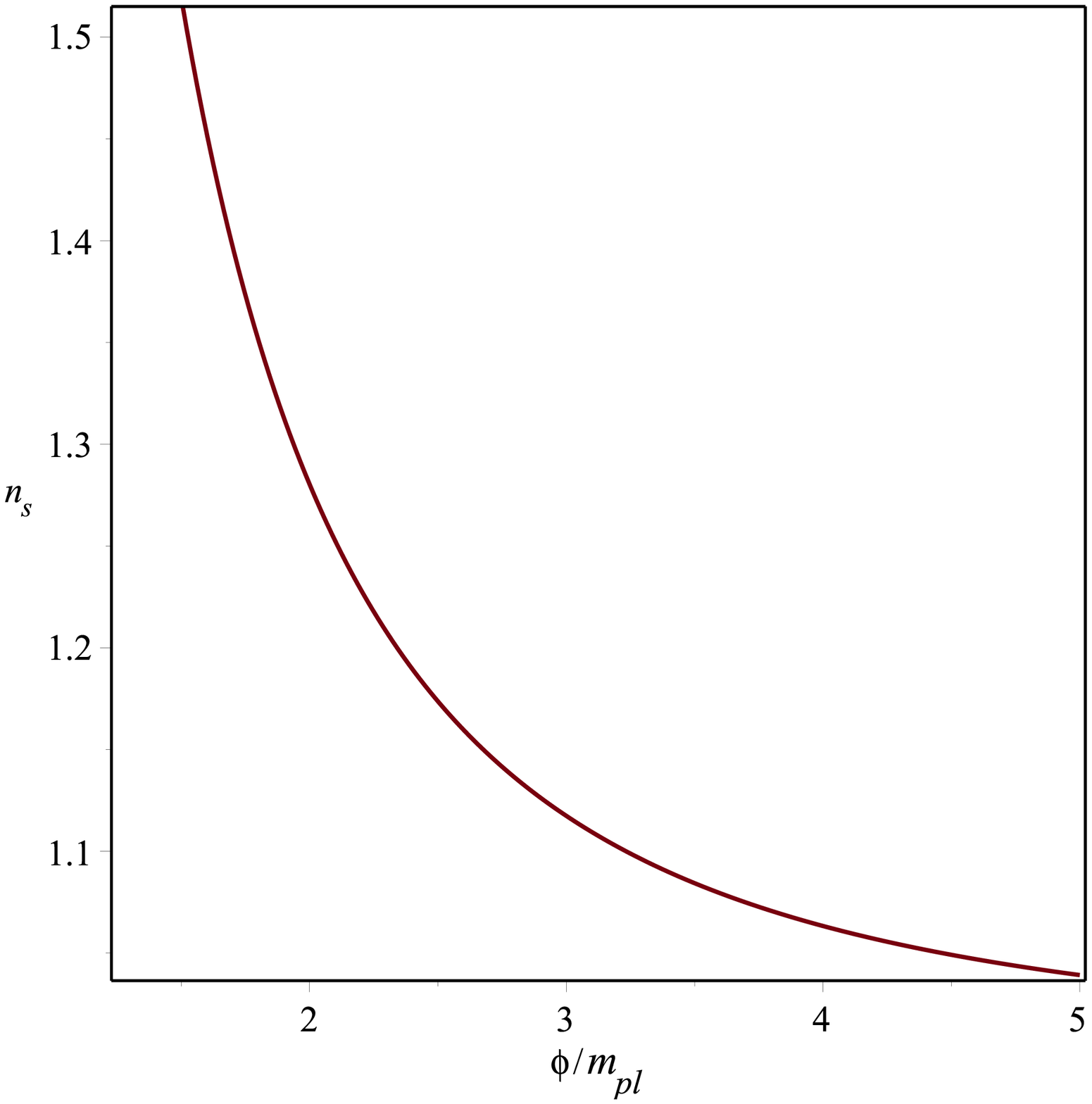} \vspace{7cm}\includegraphics{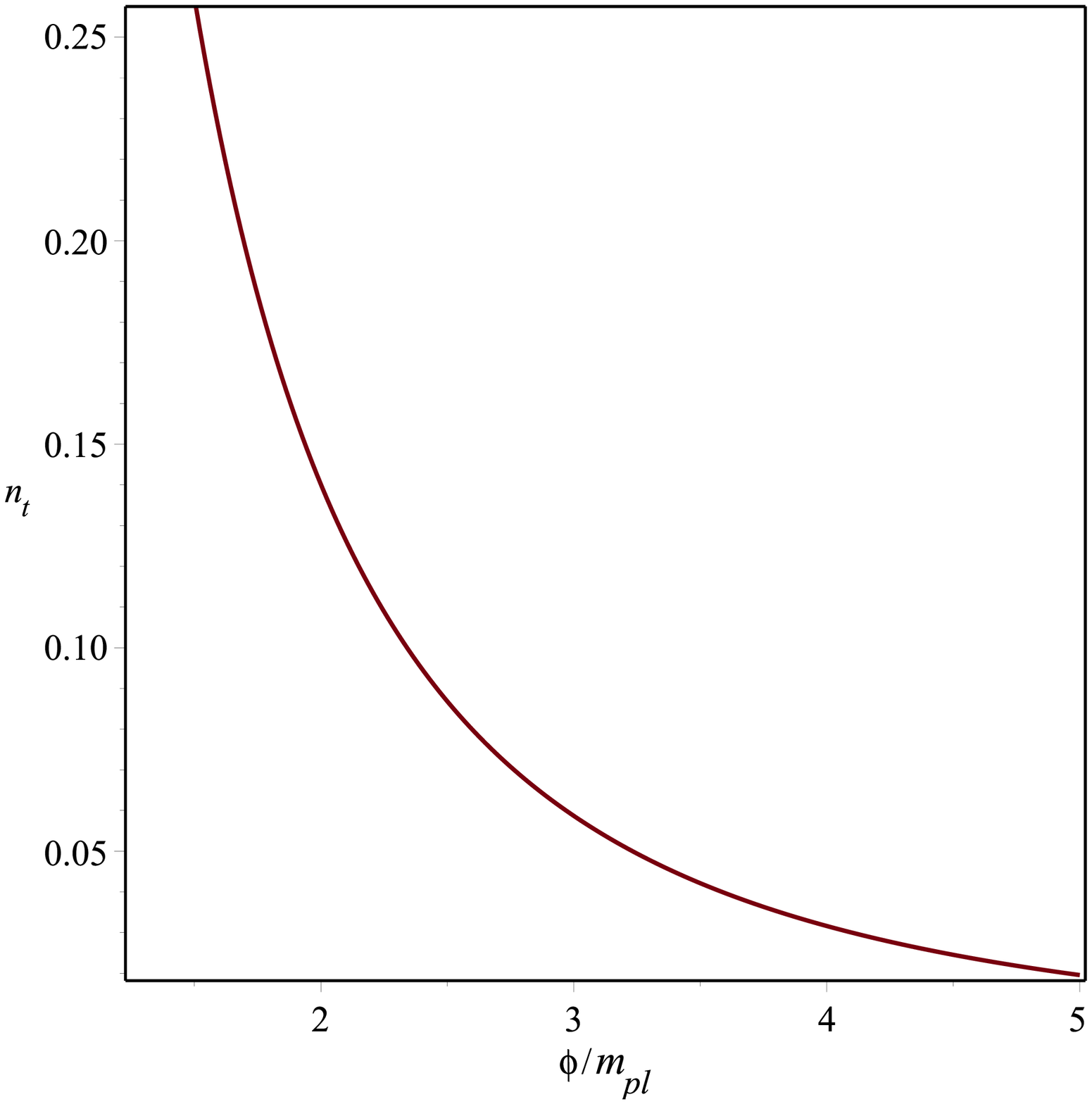}
\end{center}
\caption{\small {Variation of the scalar spectral index (left panel) and tensor spectral index
(right panel) versus $\phi/m_{pl}$  for quadratic potential in G-super-inflation.}}
\end{figure}

Figure 3 (right panel) shows the variation of the tensor spectral
index $n_{T}$ versus the scalar field. Obviously, in the presence of
Galileon-like kinetic term, the tensor spectral index is positive.
So, the tensor spectrum can be blue tilted in G-super-inflation
regime within this setup. The left panel of this figure shows the
variation of the scalar spectral index
$n_{s}$ versus the scalar field. We see that the scalar spectral index
is blue tilted in this case. So, both scalar and tensor modes tilt toward
realization of the blue spectrum in this setup. In this case
the equation of state parameter and the sound speed squared are
given by
\begin{equation}
\label{eq51}w_{\phi}\simeq-1+\frac{V_{\phi}}{6V^{2}\gamma}\big(-{\cal{K}}+\sqrt{{\cal{K}}^{2}+4\gamma
V_{\phi}}\big)
\end{equation}
and
\begin{equation}
\label{eq52}c^{2}_{s}\simeq\frac{2-\frac{{\cal{K}}}{{\cal{K}}+\sqrt{{\cal{K}}^{2}+4\gamma
V_{\phi}}}}{3\Big(1-\frac{{\cal{K}}}{{\cal{K}}+\sqrt{{\cal{K}}^{2}+4\gamma
V_{\phi}}}\Big)}\,.
\end{equation}
Also in this case the ghost instability can be avoided if the following condition is
satisfied
\begin{equation}
\label{eq53}{\cal{U}}\simeq3+\frac{2\bigg[(\frac{\dot{\phi}}{H})^{2}\bigg({\cal{K}}-4\gamma
V(\frac{\dot{\phi}}{H})+\frac{2\gamma_{\phi}V}{3}(\frac{\dot{\phi}}{H})^{2}\bigg)-6\bigg]}{3\bigg(1+\frac{\gamma
V}{6}(\frac{\dot{\phi}}{H})^{3}\bigg)^{2}}>0
\end{equation}

\begin{figure}
\begin{center}\includegraphics{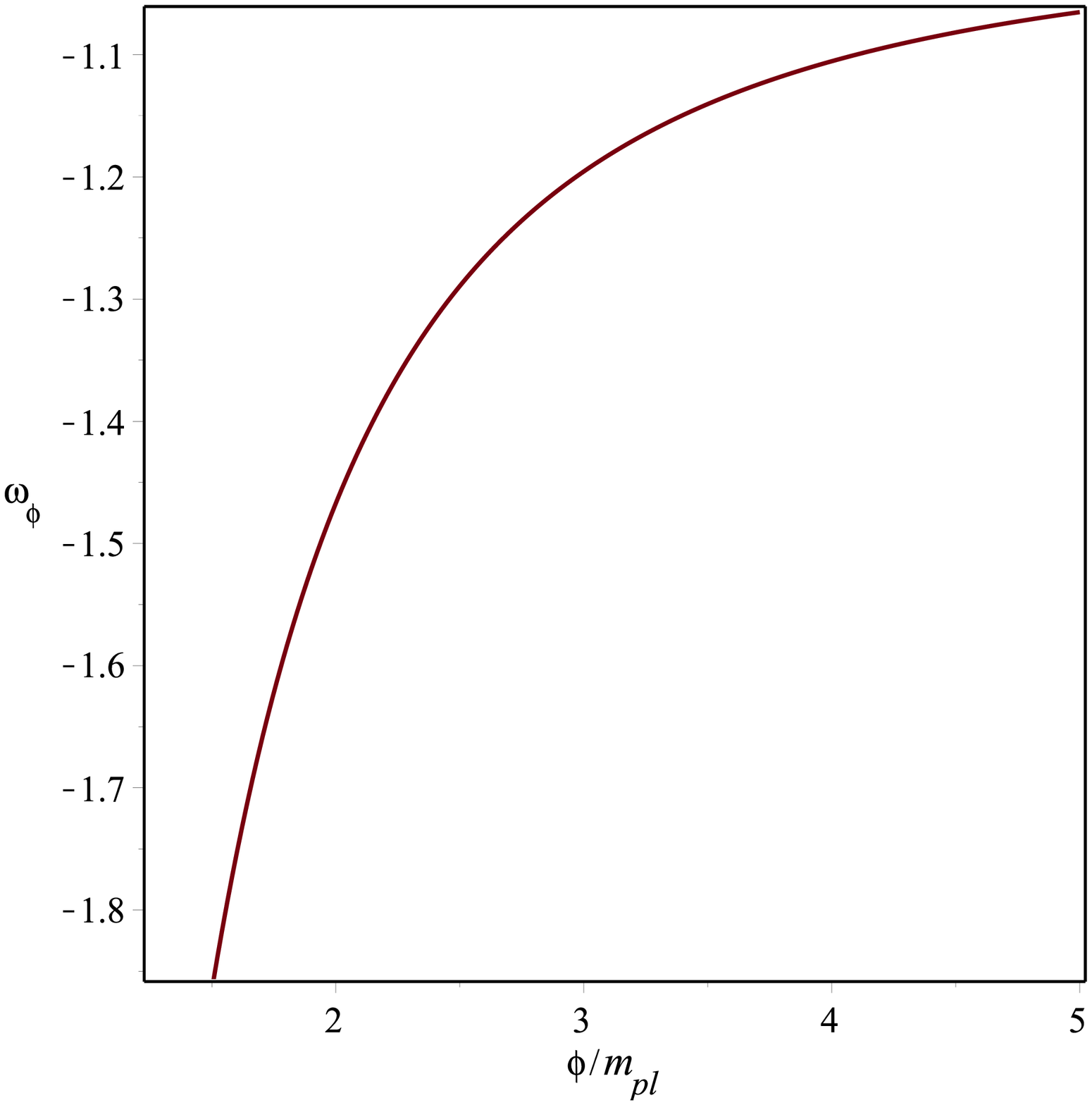} \vspace{7cm}\includegraphics{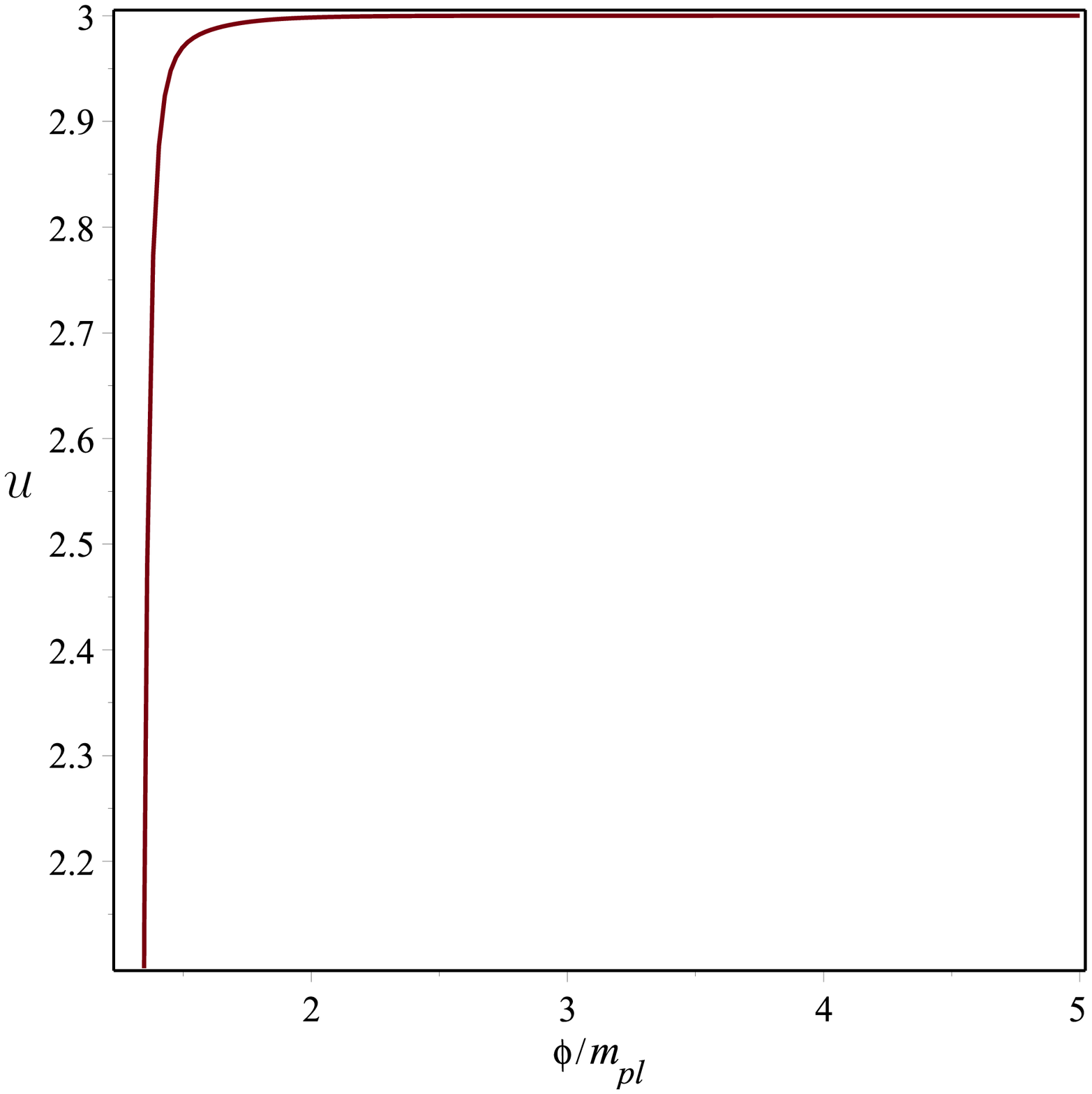}
\end{center}
\caption{\small {Variation of the equation of state parameter (left panel) and ghost instability
parameter $({\cal{U}})$ (right panel) versus $\phi/m_{pl}$  for
quadratic potential in G-super-inflation.}}
\end{figure}
Figure 4 (left panel) shows the variation of the equation of
state parameter versus the scalar field. In the presence of
Galileon-like kinetic term, the negative value of this parameter
implies super inflation period. Figure 4 (right panel) shows
the variation of instability parameter $({\cal{U}})$ versus the scalar
field. Enhancing the contribution of the kinetic energy in the action that contains non-canonical
higher order kinetic term leads to avoidance of ghost instability in this setup.

Finally, let us consider a class of generalized G-inflation model
where the potential is given by power low inflation. This setup
corresponds to the action (1) with terms as given by equations (5)-(7)
and also ${\cal{K}}(\phi)=-\phi^{1/n}$, $\gamma=-\alpha\phi^{n}$ with
$n=4$. Therefore, the condition to have rapid roll expansion in
super-inflation stage is provided by
\begin{equation}
\label{eq54}{\cal{K}}X<V(\phi),\,\,\, \mid\xi
H\phi\dot{\phi}\mid<V(\phi),\,\,\,\mid\gamma H
\dot{\phi}\mid<V(\phi),\,\,\,{\cal{K(\phi)}}\ll0\,.
\end{equation}
Using these conditions to calculate $(\frac{\dot{\phi}}{H})$, the spectral index of
perturbations can be determined by the following expression
\begin{eqnarray}
\label{eq55}n_{s}\simeq1-6\epsilon-\frac{g
F}{V\bigg({\cal{K}}+(\frac{1}{\mu^{2}})U+
\sqrt{({\cal{K}}+(\frac{1}{\mu^{2}})U)^{2}+4\gamma
F}\bigg)}\bigg[\frac{\gamma_{\phi}}{\gamma}-\frac{2({\cal{K}}_{\phi}+\frac{1}{\mu^{2}U_{\phi}})}{{\cal{K}}+
\frac{1}{\mu^{2}}U}\bigg]
\\\nonumber
+3\frac{gF_{\phi}}{V\bigg({\cal{K}}+(\frac{1}{\mu^{2}})U+
\sqrt{({\cal{K}}+(\frac{1}{\mu^{2}})U)^{2}+4\gamma F}\bigg)}
\end{eqnarray}
where $g=(m^{2}_{pl}+\xi\phi^{2})$, $F$ and $U$ are given by (18) and
\begin{equation}
\label{eq56}n_{t}\simeq-2\epsilon\simeq\frac{-2\xi\phi
F}{V\bigg({\cal{K}}+(\frac{1}{\mu^{2}})U+
\sqrt{({\cal{K}}+(\frac{1}{\mu^{2}})U)^{2}+4\gamma
F}\bigg)}\Big(1+\frac{\phi F}{V}\Big)
\end{equation}
Furthermore, we can calculate the equation of state parameter for this case as
\begin{equation}
\label{eq57}\omega_{\phi}\simeq-1-\frac{\dot{H}}{H^{2}}=-1+\frac{2\xi\phi
F}{3V\bigg({\cal{K}}+(\frac{1}{\mu^{2}})U+
\sqrt{({\cal{K}}+(\frac{1}{\mu^{2}})U)^{2}+4\gamma F}\bigg)}\,.
\end{equation}
To proceed further, we consider the following functions for this generalized G-inflation
model during the super-inflation stage
\begin{equation}
\label{eq58}{\cal{K}}=-\phi^{\frac{1}{n}},\,\,\,
\gamma=\alpha\phi^{\frac{1}{n}},\,\,\,\
V(\phi)=\phi^{4},\,\,\,\xi\simeq10^{4}.
\end{equation}
where $\alpha<1$ and in numerical analysis we set $n=4$.
\begin{figure}
\begin{center}\includegraphics{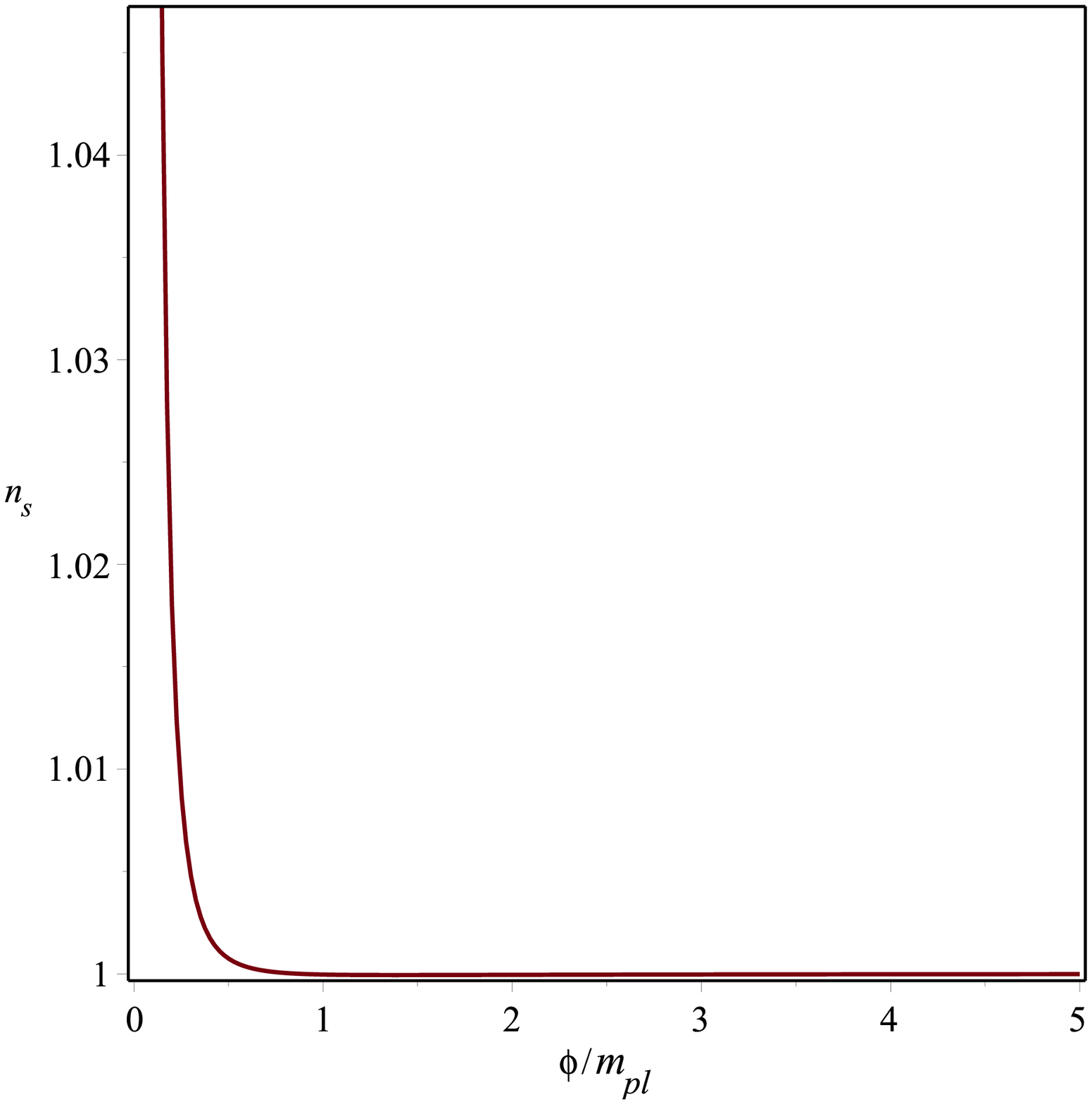} \vspace{4.5cm}\includegraphics{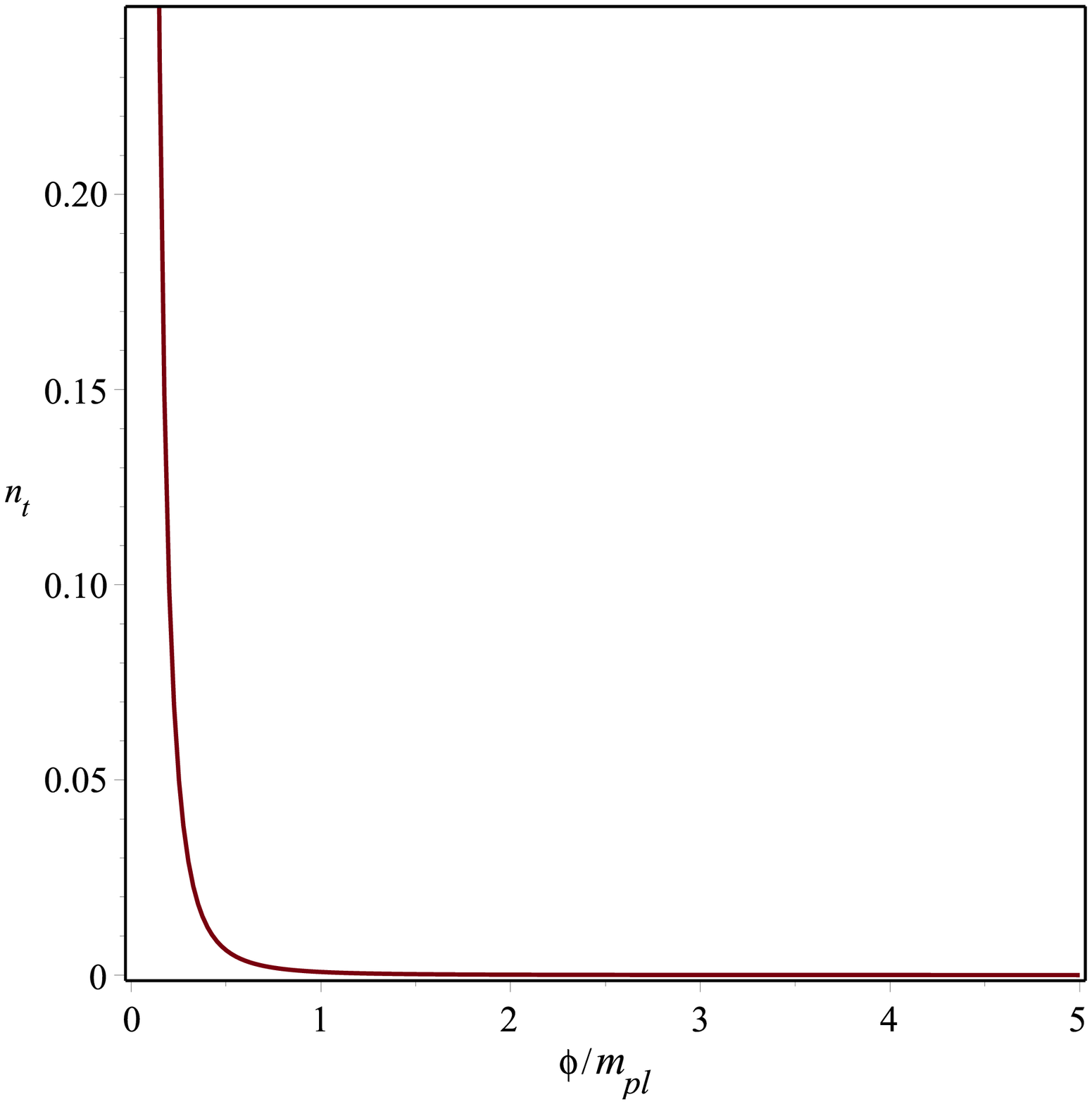}
\end{center}
\caption{\small {Variation of the scalar spectral index
(left panel) and tensor spectral index
(right panel) versus $\phi/m_{pl}$  for quadratic potential in generalized G-super-inflation.}}
\end{figure}
Figure 5 (right panel) shows the variation of the tensor spectral
index $n_{T}$ versus the scalar field. In the presence of
Galileon-like kinetic term the tensor spectral index is positive.
So, the tensor spectrum can be blue in super-inflation regime in this case.
The left panel of this figure shows the variation of the scalar spectral index
$n_{s}$ versus the scalar field. The spectral index is larger than unity and is
blue tilted in this case.
\begin{figure}
\begin{center}\includegraphics{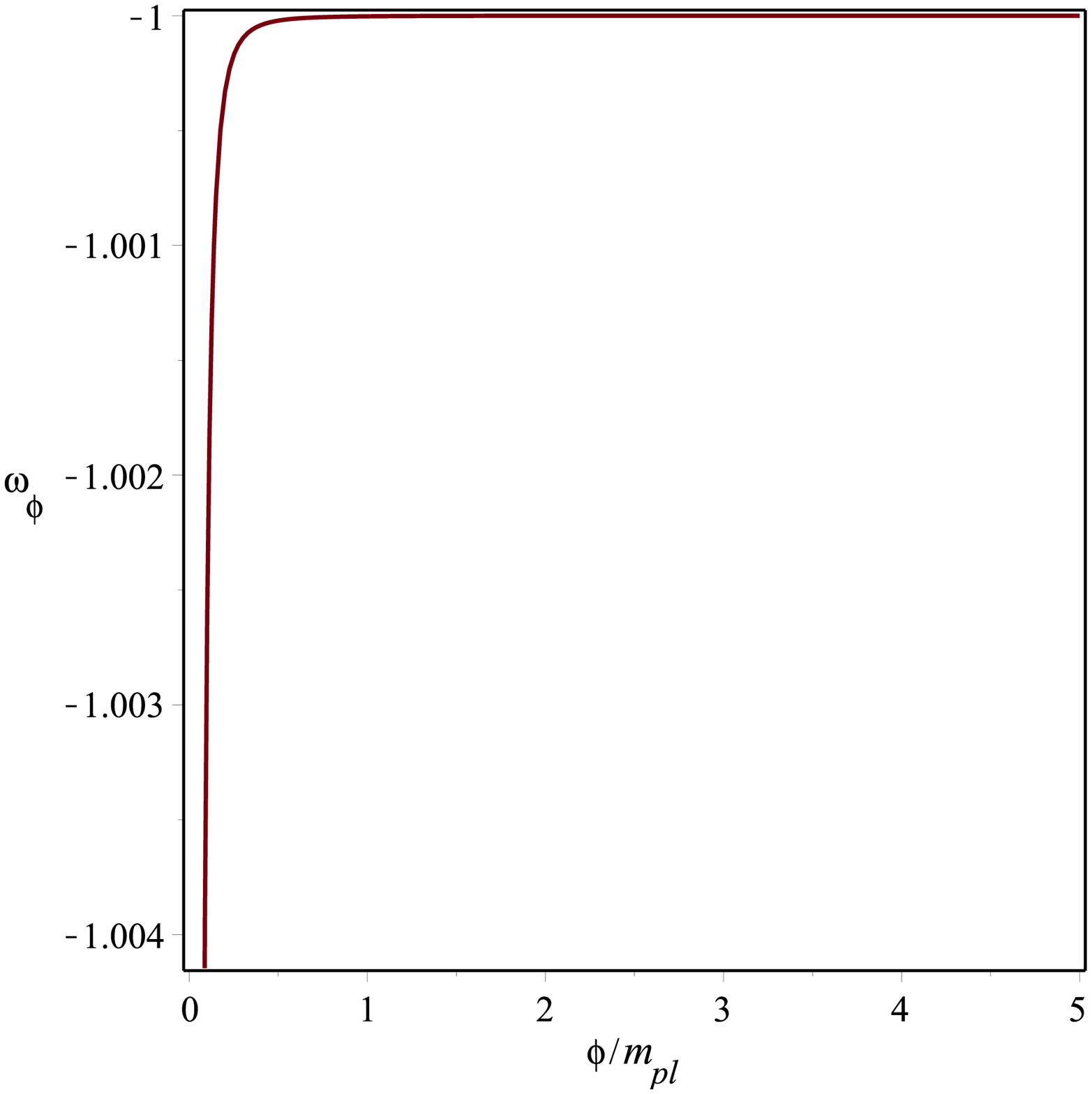} \vspace{7cm}\includegraphics{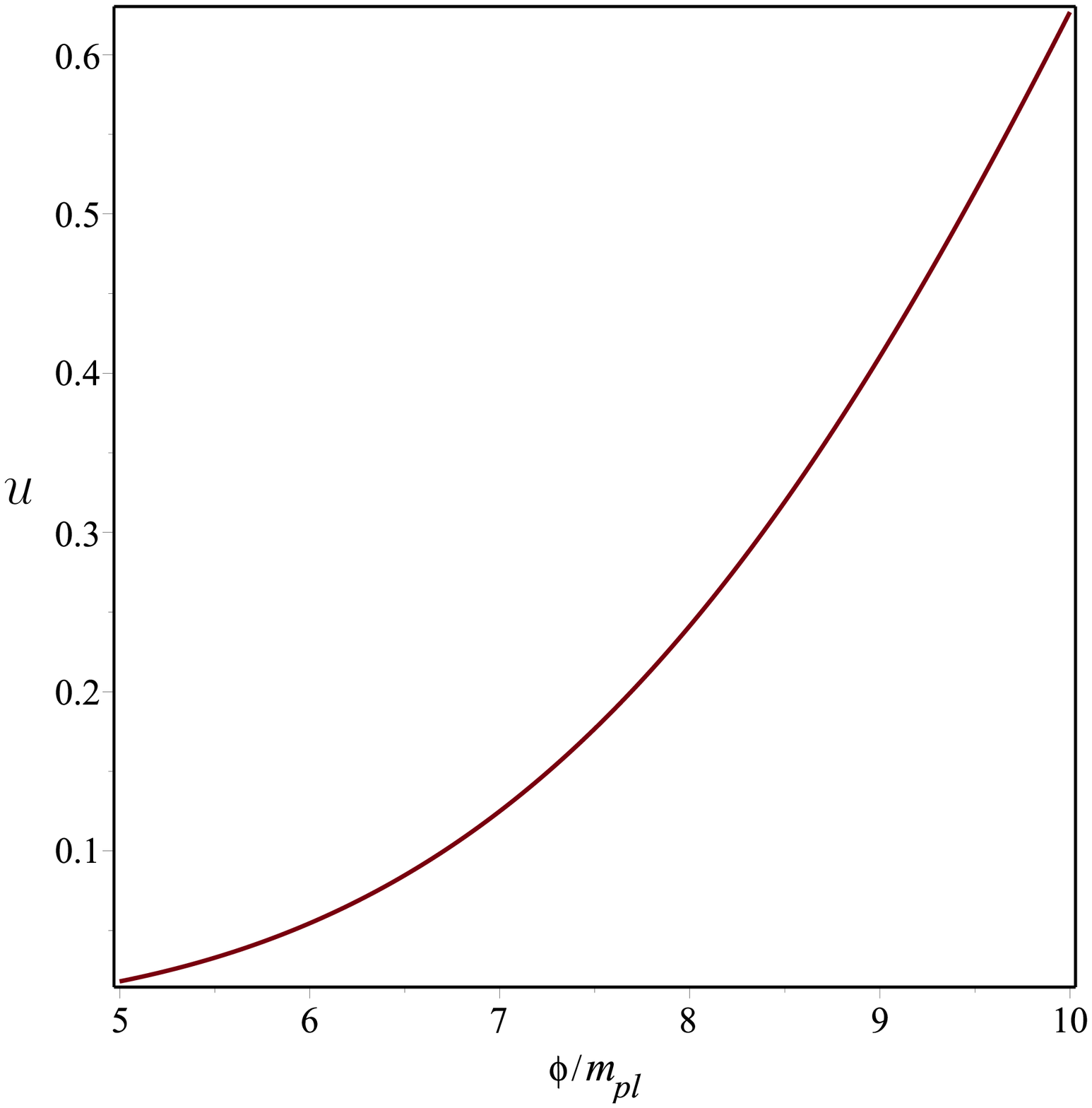}
\end{center}
\caption{\small {Variation of the equation of state parameter (left panel)
and ghost instability
parameter ${\cal{U}}$ (right panel) versus $\phi/m_{pl}$  for
quadratic potential in generalized G-super-inflation.}}
\end{figure}

Figure 6 (left panel) shows the variation of the equation of
state parameter versus the scalar field. In the presence of
Galileon-like kinetic term with non-minimal coupling between the
scalar field and gravity and the non-minimal derivative
coupling of the scalar field with Einstein tensor, the
negative value of this parameter (equation of state parameter) implies the existence of super-inflation
period in this setup. The right panel of this figure shows the variation of the instability
parameter ${\cal{U}}$ versus the scalar field. Once again, enhancing the kinetic
energy in the action that contains non-canonical
higher order kinetic terms leads to avoidance of ghost instability in this setup.

\section{Conclusion}
We have considered a generalized G-inflation model which
contains non-canonical and Galileon-like kinetic terms.
We have studied the early time cosmological dynamics of this setup
during the super-inflation period. We have calculated the
inflationary parameters and the primordial density and tensor perturbations
with details. The analysis shows that the scalar spectral index in
this setup can be larger than unity and the tensor spectral index is positive.
Therefore, scalar and tensor modes can be blue-tilted in this setup.
We have shown that presence of the Galileon-like term leads to modified
k-inflation with possibility to avoid ghost instabilities. Since the tensor
power spectrum depends only on the Hubble expansion rate, we can
conclude that possibility of having blue spectrum of primordial
perturbations is provided when the Hubble rate increases rapidly
while the null energy condition is violated. We note that in Ref. [27],
where the origin of an effective quantum-to-classical transition for perturbations is clearly explained,
it has been shown that breaking of the null and weak energy conditions occurs only near the first Hubble
radius crossing during inflation, and it is not responsible for approximate constancy of perturbations in
the strongly squeezed regime after that. Thus, the blue spectrum of gravitational wave perturbations is due
to the specific background behavior ($\dot H >0$) only, and not specifically due to the breaking of energy conditions
that occurs for any spectrum of created perturbations. As has been indicated in [28], squeezed
states as negative energy fluxes with negative pressure are
responsible for this condition. Another possible effect of
super-inflationary phase realizing blue spectrum can be
provided by modified Friedmann equation to incorporate higher
derivative extension of the Einstein's GR in the action (see [11,29-33]).
In summary, in addition to other approaches such as loop quantum cosmological considerations (see also [34] for another approach), it
is possible also to realize blue spectrum of perturbation by taking a super-inflation stage into account.
In this manner and via a generalized Galileon-like kinetic term in the action, it is possible to
avoid ghost instability that are usually present in k-inflation models.


\begin{thebibliography}{11}

\bibitem{Lin82}
A. A. Starobinsky, Phys. Lett. B \textbf{91}, 99 (1980)\\
A. H. Guth, Phys. Rev. D \textbf{23}, 347 (1981)\\
K. Sato, MNRAS \textbf{195}, 467 (1981)\\
A. D. Linde, Phys. Lett. B \textbf{108}, 389 (1982)\\
A. Albrecht and P. J. Steinhardt, Phys. Rev. Lett. \textbf{48}, 1220 (1982)\\
S. W. Hawking and I. G. Moss, Phys. Lett. B \textbf{110}, 35 (1982)\\
A. D. Linde, Phys. Lett. B \textbf{129}, 177 (1983)\\
A. R. Liddle and D. H. Lyth, cosmological Inflation and Large-Scale
Structure, Cambridge University Press, (2000).


\bibitem{Ru82}
A. A. Starobinsky, JETP Lett. \textbf{30}, 682  (1979)\\
V. F. Mukhanov and G. V. Chibisov, JETP Lett.  \textbf{33}, 532  (1981)\\
V. A. Rubakov, M. V. Sazhin and A. V. Veryaskin, Phys. Lett. B \textbf{115},  189 (1982)\\
R. Fabbri and M. D. Pollock, Phys. Lett. B \textbf{125}, 445 (1983)\\
L. F. Abbott and M. B. Wise, Nucl. Phys. B \textbf{244}, 541 (1984)\\
A. A. Starobinsky, Sov. Astron. Lett. \textbf{11}, 133 (1985)\\
L. F. Abbott and D. D. Harari, Nucl. Phys. B \textbf{264}, 487 (1986).

\bibitem{Ki20}
W. H. Kinney, A. Melchiorri and A. Riotto, Phys. Rev. D \textbf{63}, 023505 (2001)\\
W. H. Kinney, E. W. Kolb, A. Melchiorri and A. Riotto, Phys. Rev. D \textbf{78}, 087302 (2008)\\
S. Dodelson, W. H. Kinney and E. W. Kolb, Phys. Rev. D \textbf{56}, 3207 (1997)\\
J. Martin, C. Ringeval, R. Trotta and V. Vennin, JCAP \textbf{1403}, 039 (2014).

\bibitem{MK97}
M. Kamionkowski, A. Kosowsky and A. Stebbins, Phys. Rev. D \textbf{55}, 7368 (1997)\\
M. Zaldarriaga and U. Seljak, Phy. Rev. D \textbf{55}, 1830 (1997).

\bibitem{AB16}
B. P. Abbott et al. \emph {LIGO Scientific and Virgo
Collaborations}, Phys. Rev. Lett. \textbf{116}, 6, 061102 (2016)
[arXiv:1602.03837[gr-qc]]\\
B. Abbott et al. [ALLEGRO Collaboration], [arXiv:gr-qc/0703068].

\bibitem{YC16}
Yong Cai, Y. -T. Wang, and Y. -S. Piao \emph {Propagating speed of
primordial gravitational waves and inflation}, Phys. Rev. D \textbf{94}, 043002 (2016)\\
Y.-F. Cai, J.-O. Gong, S. Pi, E. N.Saridakis and S.-Y. Wu, Nucl. Phys. B \textbf{900}, 517 (2015)\\
Y.-F. Cai, [arXiv:1405.1369].

\bibitem{MB05}
M. Baldi, F. Finelli and S. Matarrese, Phys. Rev. D \textbf{72}, 083504 (2005).

\bibitem{YS04}
Y. S. Piao and Y. Z. Zhang, Phys. Rev. D. \textbf{70}, 063513 (2004)\\
Z. G. Liu, Z. K. Guo and Y. S. Piao, Eur. Phys. J. C \textbf{74}, 3006 (2014)\\
Yong Cai, Y. T. Wang and Y. S. Piao, Phys. Rev. D \textbf{93}, 063005 (2016).

\bibitem{DM06}
D. Mulryne and N. Nunes, Phy. Rev. D \textbf{74}, 083507 (2006)\\
J. Grain, A. Barrau, T. Cailleteau and J. Mielczarek, Phys. Rev. D \textbf{82}, 123520 (2010).

\bibitem{MB02}
M. Bojowald, Phys. Rev. Lett. \textbf{89}, 261301 (2002)\\
M. Bojowald and K. Vandersloot, Phys. Rev. D \textbf{67},
124023 (2003)\\
X. -J. Yue and J. -Y. Zhu, Phys. Rev. D \textbf{87}, 063518 (2013).

\bibitem{As06}
A. Ashtekar, T. Pawlowski and P. Singh, Phys. Rev. D \textbf{74}, 084003 (2006)\\
J. Ribassin, E. Huguet and K. Ganga, [arXiv:gr-qc/1111.4661].

\bibitem{Bi13}
T. Biswas and A. Mazumdar [arXiv:hep-th/1304.3648].

\bibitem{TK10}
T. Kobayashi, M. Yamaguchi and J. Yokoyama, Phys. Rev. Lett. \textbf{105}, 231302 (2010).

\bibitem{CB11}
C. Burrage, C. de Rham, D. Seery and A. J. Tolley, JCAP \textbf{1101}, 014 (2011).

\bibitem{KK11}
K. Kamada, T. Kobayashi, M. Yamaguchi and J. Yokoyama, Phy. Rev. D. \textbf{83}, 083515 (2011).

\bibitem{TK11}
T. Kobayashi, M. Yamaguchi, J. Yokoyama, Prog. Theor. Phys. \textbf{126}, 511 (2011).

\bibitem{TK13}
T. Kobayashi, N. Tanahashi, M. Yamaguchi, Phys. Rev. D. \textbf{88}, 083504 (2013).

\bibitem{AD11}
A. De Felice and S. Tsujikawa, Phys. Rev. D. \textbf{84}, 083504 (2011)\\
A. De Felice and S. Tsujikawa, JCAP \textbf{1203}, 025 (2012).

\bibitem{Ho74}
G. W. Horndeski, Int. J. Theor. Phys. \textbf{10}, 363 (1974)\\
A. De Felice and S. Tsujikawa, JCAP \textbf{03}, 030 (2013)\\
C. Burrage, C. de Rham, D. Seery and A. J. Tolley, JCAP \textbf{1101}, 014 (2011).

\bibitem{Wa84}
R. M. Wald, \emph{General Relativity}, The University of Chicago Press, Chicago, 1984\\
R. L. Arnowitt, S. Deser and C. W. Misner, Phys. Rev. D \textbf{117}, 1595 (1960).

\bibitem{Ar04}
R. L. Arnowitt, S. Deser and C. W. Misner, \emph{Gravitation: an introduction to current research},
Louis Witten ed. (Wiley 1962), chapter 7, pp 227--265, [arXiv:gr-qc/0405109]\\
J. Bardeen, Phys. Rev. D  \textbf{22}, 1882 (1980).

\bibitem{Mu92}
V. F. Mukhanov, H. A. Feldman and R. H. Brandenberger, Phys. Rept. \textbf{215}, 203 (1992).

\bibitem{Be95}
E. Bertschinger, [arXiv:astro-ph/9503125].

\bibitem{De11}
A. De Felice and S. Tsujikawa, JCAP \textbf{1104}, 029 (2011).

\bibitem{Ka12}
K. Kamada, T. Kobayashi, T. Takahashi, M. Yamaguchi and Jun'ichi Yokoyama, [arXiv:1203.4059].

\bibitem{Ma11}
A. Mazumdar and J. Rocher, Phys. Rept. \textbf{497}, 85 (2011).

\bibitem{Pol96}
D. Polarski and A. A Starobinsky, Class. Quant. Grav. \textbf{13}, 377 (1996).

\bibitem{Fo97}
L. H. Ford \emph{Quantum Field Theory In Curved Spacetime}, [arXiv:gr-qc/9707062].

\bibitem{TB06}
T. Biswas, A. Mazumdar and W. Siegel, JCAP \textbf{0603}, 009 (2006).

\bibitem{TB10}
T. Biswas, T. Koivisto and A. Mazumdar, JCAP \textbf{1011}, 008 (2010).

\bibitem{TB12}
T. Biswas, A. S. Koshelev, A. Mazumdar and S. Y. Vernov, JCAP \textbf{1208} 024 (2012).

\bibitem{SY89}
A. S. Koshelev and S. Y. Vernov, Phys. Part. Nucl. \textbf{43}, 666 (2012).

\bibitem{AK12} A. S. Koshelev, [arXiv:astro-ph.CO/1302.2140]\\
A. S. Koshelev, Rom. J. Phys. \textbf{57}, 894 (2012).

\bibitem{GD15}
G. Domènech and M. Sasaki, JCAP 1504 (2015) 022.

\end{thebibliography}
\end{document}